\documentclass[aps,prc,reprint,groupedaddress]{revtex4-2}
\usepackage{amsmath}
\usepackage{amssymb}
\usepackage{wallpaper}
\usepackage{graphicx}
\usepackage{dcolumn}
\usepackage{bm}
\usepackage{color}
\usepackage{multirow}
\usepackage{CJK}
\usepackage[T1]{fontenc}
\usepackage{mathptmx}
\usepackage{tabularx}
\usepackage{ulem}
\usepackage[colorlinks,linkcolor=blue,anchorcolor=blue,citecolor=blue]{hyperref}

\begin{document}
\begin{CJK*}{UTF8}{gbsn}
		\renewcommand\arraystretch{1.2}
		
\title{Probing the structure of pygmy dipole resonance with its $\gamma$ decay}
		
\author{W.-L. Lv
} 
\address{Frontiers Science Center for Rare isotopes, Lanzhou University, Lanzhou 730000, China}
\address{School of Nuclear Science and Technology, Lanzhou University, Lanzhou 730000, China}
\author{Y.-F. Niu
}\email{niuyf@sjtu.edu.cn} 
\address{School of Physics and Astronomy, Shanghai Jiao Tong University, 
         Key Laboratory for Particle Astrophysics and Cosmology (MoE), Shanghai 200240, China}
\address{Shanghai Key Laboratory for Particle Physics and Cosmology, Shanghai 200240, China}
\address{Frontiers Science Center for Rare isotopes, Lanzhou University, Lanzhou 730000, China}
\address{School of Nuclear Science and Technology, Lanzhou University, Lanzhou 730000, China}
\author{G. Col\`{o}}
\address{Dipartimento di Fisica, Universit\`{a} degli Studi di Milano, via Celoria 16, I-20133 Milano, Italy}
\address{INFN, Sezione di Milano, via Celoria 16, I-20133 Milano, Italy}

\date{\today}

\begin{abstract}
The isospin properties and the collectivity of the pygmy dipole resonance (PDR) are long-standing open questions in nuclear structure studies. To answer these questions,  
the $\gamma$-decay  of PDR states  in $^{208}$Pb  to the low-lying $2_{1}^{+}$ state is investigated 
using the Skyrme particle-vibration coupling (PVC) model.
It is found that the $E1$ $\gamma$ decay from the PDR states to the low-lying $2_{1}^+$ state
is strongly suppressed compared to that from the isovector giant dipole resonance (IVGDR),
which reveals the predominantly isoscalar character of PDR.
A detailed decomposition of the decay diagrams 
of the amplitudes of the processes contributing to the decay demonstrates the 
non-negligible presence of the 
1 particle-1 hole configurations coupled to the $2_1^{+}$ phonon in the PDR wave function.
Furthermore, we give a quantitative way to identify the components of 
complex-configurations in the wave function, and it is found that such component in PDR is smaller than
that in IVGDR and much smaller than that in isoscalar giant quadrupole resonance (ISGQR). 
\end{abstract}

\maketitle
\end{CJK*}

\section{Introduction}
\label{secIntro}




Nuclear collective phenomena have been a long-standing topic of great interest, 
as they elucidate the fundamental properties of nuclear matter \cite{Bortignon1998,Harakeh2001}. 
Among them, the pygmy dipole resonance (PDR),
a concentration of electric dipole strength around the neutron emission threshold,
has been and still is extensively investigated \cite{Paar2007,Savran2013,Bracco2019,NeumannCosel2019,Lanza2023}.
In medium-mass and heavy nuclei, 
it is often interpreted as an oscillation of a neutron skin against the isospin saturated core.
Guided by this picture, PDR data are used to constrain the density dependence of nuclear symmetry energy,
and consequently the thickness of neutron skin 
\cite{Piekarewicz2006,Klimkiewicz2007,Carbone2010,Inakura2011,Papakonstantinou2015,Li2021}.
The presence of PDR can strongly enhance the neutron capture rate involved in cosmic nucleosynthesis
\cite{Goriely1998,Litvinova2009,Tsoneva2015,Tonchev2017}.
For these reasons, PDR has been widely measured in stable and unstable nuclei with neutron excess
\cite{Adrich2005,Klimkiewicz2007a,Wieland2009,Rossi2013,Angell2012,Goddard2013,Massarczyk2014,Krumbholz2015,
Savran2018,Renstroem2018,Martorana2018,Spieker2020,Bassauer2020,Crespi2021,Weinert2021}.
However, there are still open questions regarding the isospin character and the collectivity. 
In fact, it remains debated whether the low-energy $E1$ strength is a genuine collective resonance 
or rather a fragmented predominantly single-particle excitation mode.


The isospin character of dipole states in the PDR region is complex.
Below the neutron emission threshold,
the low-energy part of the PDR can be populated by both isoscalar and isovector probes,
whereas the higher-energy part is primarily excited by the electromagnetic interaction
\cite{Savran2006,Savran2011,Derya2014,Crespi2014,Pellegri2014,Krzysiek2016,Crespi2015,Negi2016}.
This phenomenon is known as isospin splitting. In contrast, 
above the threshold, a recent experiment on $^{68}$Ni \cite{Martorana2018} has demonstrated that
in addition to the isovector probe (virtual photon) \cite{Wieland2009,Rossi2013},
the dipole states at the tail of GDR can also be excited by an isoscalar probe (isoscalar $^{12}$C target).
However, due to limitations in statistics and the relatively scarce energy resolution, 
more precise measurements are necessary to better prove this observation.
Theoretically, the isospin character of a state is distinguished by the relative
oscillation phase of the proton and neutron transition densities 
\cite{Tsoneva2008,Co2009,Litvinova2009a,Paar2009,Niu2009,Gambacurta2011,Lanza2011}.
The isospin character is important to obtain the maximum response strength 
under the corresponding probe.
Dipole states with different isospin characters need to be disentangled 
to provide a benchmark for theoretical models.




The collectivity of PDR is still under debate.
To investigate its collectivity microscopically, 
one must first identify the specific configurations, 
e.g., 1 particle-1 hole (1p-1h) and 2 particle-2 hole (2p-2h),
that contribute to the wave function 
and then analyze their coherence \cite{Lanza2023}.
At 1p-1h level, from the analyses of quasiparticle random phase approximation (QRPA),
the isoscalar response of PDR states exhibits clear collectivity 
due to the coherent sum of the single-neutron transition amplitudes, 
whereas the isovector response does not \cite{RocaMaza2012,Vretenar2012}. 
To better reproduce the dipole states below the neutron emission threshold,
one needs to go beyond QRPA and consider 2p-2h configurations \cite{Hartmann2004},
like in the second random phase approximation (SRPA) \cite{Gambacurta2011}, 
in large-scale shell model (LSSM) \cite{Brown2000},
and in the quasiparticle-phonon model (QPM) \cite{Tsoneva2008}.
In $^{208}$Pb \cite{Spieker2020}, 
by analyzing the differential cross sections with the distorted-wave Born-approximation (DWBA),
a non-collective 1p-1h nature of these states is suggested,
which is also supported by the LSSM and QPM calculations.
At the same time, they assign a substantial role to complex 2p-2h or two-phonon components in the wave function above the threshold, 
although in this near-threshold low-energy region they do not reproduce the measured $E1$ strength distribution.
Identifying the 2p-2h configuration is important,
because coupling to complex configurations can redistribute 
and fragment the dipole strength and thereby modify the coherence.


In the past decades, $\gamma$ decay has been proved to be a powerful tool
to study the structure of nuclear resonances \cite{Bracco2019}.
Besides the $\gamma$-decay to the ground state,
$\gamma$ decay to low-lying states is a significant source of information \cite{Isaak2013,Loeher2016,Muescher2020,Papst2020}.
The $\gamma$-decay width of nuclear resonances to low-lying states is small.
Hence the experimental data are rare.
For the giant quadrupole resonance (GQR), its decay to $3_{1}^{-}$ and to the ground state 
has only been reported in Ref.~\cite{Beene1989} for $^{208}$Pb.
Very recently, the $\gamma$-decay from dipole resonances to the $2_{1}^+$ state has been measured
by nuclear resonance fluorescence (NRF) experiments with High Intensity $\gamma$-ray Source (HI$\gamma$S) \cite{Kleemann2025,Papst2025}.
In Ref. \cite{Kleemann2025}, the authors measured the $\gamma$-decay from IVGDR to $2_{1}^+$ state in $^{154}$Sm.
They successfully extracted the deformation parameters by comparing the branching ratio $\sigma_{2_{1}^{+}}/\sigma_{\rm ES}$ 
with the macroscopic geometrical model.
In Ref. \cite{Papst2025}, the authors measured $\gamma$-decay from PDR to $2_1^+$ state in $^{150}$Nd.
They found a strong deviation from the Porter-Thomas distribution,
which would alter the photon strength functions (PSFs) used in statistical models \cite{TALYS2023}.
These findings confirm the $\gamma$-decay to low-lying states as an observable sensitive to the structure of resonance
and its importance to nuclear astrophysics.
More measurements will be carried out at 
the Laboratori Nazionali di Legnaro (LNL) \cite{LNL2023}
and Shanghai Synchrotron Radiation Facility (SSRF) \cite{Chen2023}.

Theoretically, with a phenomenological nuclear Hamiltonian,
the nuclear field theory (NFT) \cite{Bortignon1984,Bes1986},
the extended theory of finite Fermi systems (ETFFS) \cite{Speth1985}, 
and the quasiparticle phonon model (QPM) \cite{Voronov1990,Ponomarev1992}
have been employed to study $\gamma$-decay from GRs to low-lying states.
The effects of the isospin character of the GRs \cite{Bortignon1984,Speth1985}
and of the admixture of complex configurations in their wave function \cite{Voronov1990,Ponomarev1992}
have been discussed.
Recently, the fully self-consistent treatment of $\gamma$-decay of GRs, 
within the Skyrme PVC model has become available \cite{Brenna2012}.
By comparing the $\gamma$-decay widths from the GDR to $2_{1}^{+}$ state and from the GQR to $3_1^{-}$ state in $^{208}$Pb,
a much larger weight of the low-lying phonon component in the GQR wave function is deduced \cite{Lv2021},
which demonstrates that the $\gamma$-decay to low-lying states 
is a unique probe of the resonance wave function.
This strongly motivates us to study the $\gamma$-decay of PDR,
so as to analyze its isospin character and identify the complex configurations in the wave function.

\section{Formalism}
\label{secTheo}

For the $\gamma$-decay between two vibrational states $|n_i\rangle$ and $|n_f\rangle$,
we have to consider the interplay between nuclear collective motion and individual particles.
This can be dealt with by treating the residual interaction $V$ as a perturbation, 
and considering up to the second-order correction to  $|n_i\rangle$ and $|n_f\rangle$.
Hereafter, we refer to $|n^{(k)}\rangle$ as the $k$-th order correction to $|n\rangle$.

Here we only consider the $\gamma$-decay through $E1$ transition.
The electric multipole operator reads
\begin{equation}
  Q_{1\mu}
= 
  \sum_{i=1}^{A}
  e^{\textrm{eff}}_{i}
  r_{i}  Y_{1\mu}(\hat{\bm{r}}_i),
  \label{eq:Q_lm}
\end{equation}
where the $e_i^{\rm eff}$ is the effective charge
to take into account the recoil of the nucleus, with 
$e_{\rm n}^{\rm eff} = - Z/A$ for the neutron and $e_{\rm p}^{\rm eff} = N/A$ for the proton
\cite{Bohr1998I}.

\begin{figure}[t]
\centering
\includegraphics[width=0.95\linewidth]{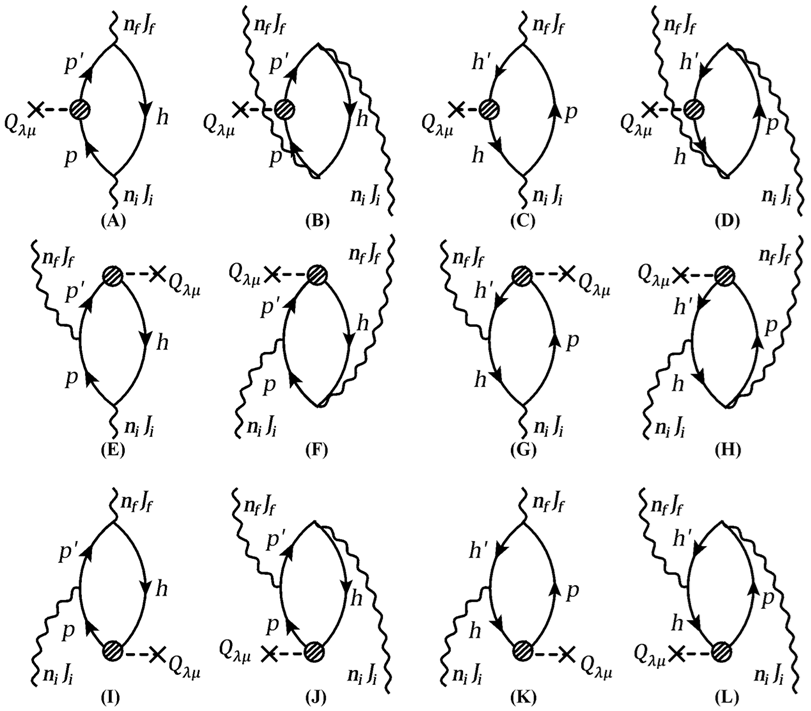}
\caption{The 12 lowest-order diagrams associated with the $\gamma$-decay between two vibrational states. 
    The circle with lines includes the contribution to $Q_{\lambda\mu}$ from nuclear polarization \cite{Bohr1998II}.}
\label{fig:fig1}
\end{figure}

\begin{figure*}[t]
    \centering
    \includegraphics[width=0.95\linewidth]{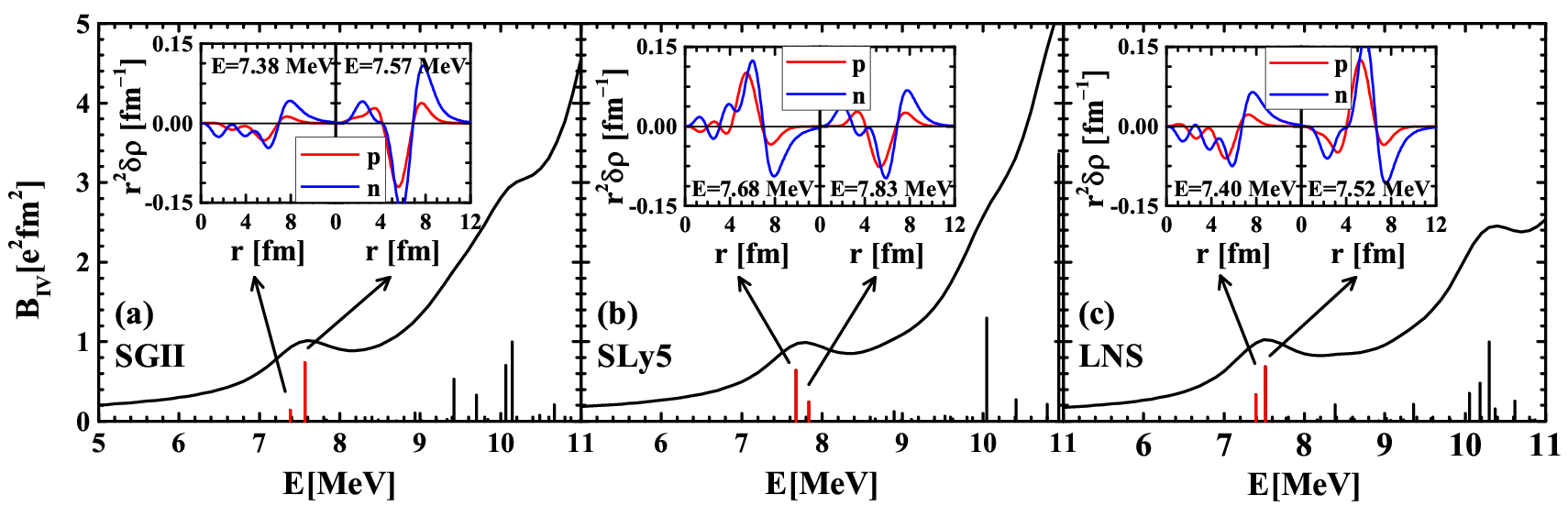}
    \caption{Discrete isovector dipole transition strengths in $^{208}$Pb for SGII (a), SLy5 (b), and LNS (c).
        The black curves are the corresponding strength functions obtained by Lorentzian smoothing with $\Gamma/2=0.5$ MeV.
        The proton (red line) and neutron (blue line) transition densities of the dipole states
        around neutron emission threshold are shown in the inset graphs.}
    \label{fig:fig2}
\end{figure*}

By considering the perturbation corrections to the initial and final wave functions,
$|n_i\rangle$ and $|n_f\rangle$, 
there are 12 terms, as sketched by the diagrams in Fig. \ref{fig:fig1}, contributing to the
transition strength
\begin{equation}
  B_{\gamma}^{fi}
= \frac{1}{2J_i +1} |\sum_{m = A, \dots, L} \langle n_f|| Q_{1} || n_i \rangle_{(m)} |^2.
\end{equation}
For the detailed expressions of the diagrams, we refer to Ref. \cite{Brenna2012}.
Diagram A - D correspond to $\langle n_f^{(1)}|| Q_{1} || n_i^{(1)} \rangle$.
These 4 diagrams are at the RPA level, 
because these two first-order corrections in $|n^{(1)}\rangle$ are respectively 
proportional to $X_{ph}$ and $Y_{ph}$ amplitudes of the RPA model.
Diagram E - H correspond to $\langle n_f^{(0)}|| Q_{1} || n_i^{(2)} \rangle$.
These diagrams include the process of initial phonon $|n_i\rangle$ scattering into $|p'h'n_f \rangle$.
Therefore, they will contribute remarkably if the initial GDR or PDR states have the configuration of 1$p$-1$h$
coupled with the final $2_{1}^{+}$ state.
Diagram I - L correspond to $\langle n_f^{(2)}|| Q_{1} || n_i^{(0)} \rangle$.
These diagrams include the process of final phonon $|n_f\rangle$ scattering into $|phn_i \rangle$.
Therefore, they will contribute remarkably if the final $2_{1}^{+}$ state has the configuration of 1$p$-1$h$
coupled with the GDR or PDR phonon.
Diagram E - L are at the PVC level,
because the two terms in $|n_i^{(2)}\rangle$ include 
the PVC vertices $\langle p'h'n_f| V |ph \rangle$ and $\langle p'h'| V |phn_i \rangle$.

\section{Numerical details}
\label{secNume}

The initial and final vibrational states are calculated with the fully self-consistent RPA method \cite{Colo2013}.
The box size for calculating the single-particle levels is 20 fm.
The energy cut-off for single-particle states involved in 1$p$-1$h$ configuration is 150 MeV.
The model space is large enough so that the double commutator energy weighted sum rule (EWSR) is satisfied.
For SLy5 \cite{Chabanat1998SLy45}, EWSR is exhausted by 100.1\%.
In this study the GDR states are selected from the RPA states with energies of
10-18 MeV and a fraction of isovector
(IV) EWSR that is larger than 5\%.
The PDR states are selected by the RPA states lying  between 6 and 8 MeV. 

\section{Results and discussions}
\label{secResu}


We study the dipole response in $^{208}$Pb with 
three energy density functionals, SGII \cite{Giai1981}, SLy5 \cite{Chabanat1998SLy45}, and LNS \cite{Cao2006LNS}.
Their slopes of the symmetry energy at saturation density $L$
span a wide range, namely
37.6 MeV, 48.2 MeV, and 61.5 MeV, respectively.
The strength distributions of the isovector dipole operator are depicted
in Fig. \ref{fig:fig2}.
In all cases,
low-lying resonance peaks are observed around the neutron emission threshold $E_{\rm th}=7.37$ MeV,
consistent with the concentration of low-energy $E1$ strength 
    around $E_{\rm th}$ reported in the $(\gamma,\gamma')$ experiment \cite{Ryezayeva2002}.
We plot the proton and neutron transition densities of the dipole states 
around $E_{\rm th}$ in the inset panels.
For all three Skyrme functionals,
the proton and neutron transition densities show a similar radial pattern 
in the nuclear interior,
while the surface region is dominated by the neutron transition density. In this respect,
the dipole states can be viewed as PDR states.
The fractions of the EWSR exhausted by the selected pygmy dipole states 
are 0.65\%, 0.78\%, and 0.81\%, respectively, for SGII, SLy5, and LNS.
These values reasonably agree with the experimental value $0.92 \pm 0.09 \%$ \cite{Poelhekken1992}.
Note that the experimental value is derived from summing the EWSR fractions of the individual states 
located within the 7-8 MeV interval, ensuring a consistent comparison with our theoretical selection.
The exhaustion of EWSR increases with $L$, which is in line with Ref. \cite{Carbone2010}.

\begin{figure}[t]
\centering
\includegraphics[width=0.8\linewidth]{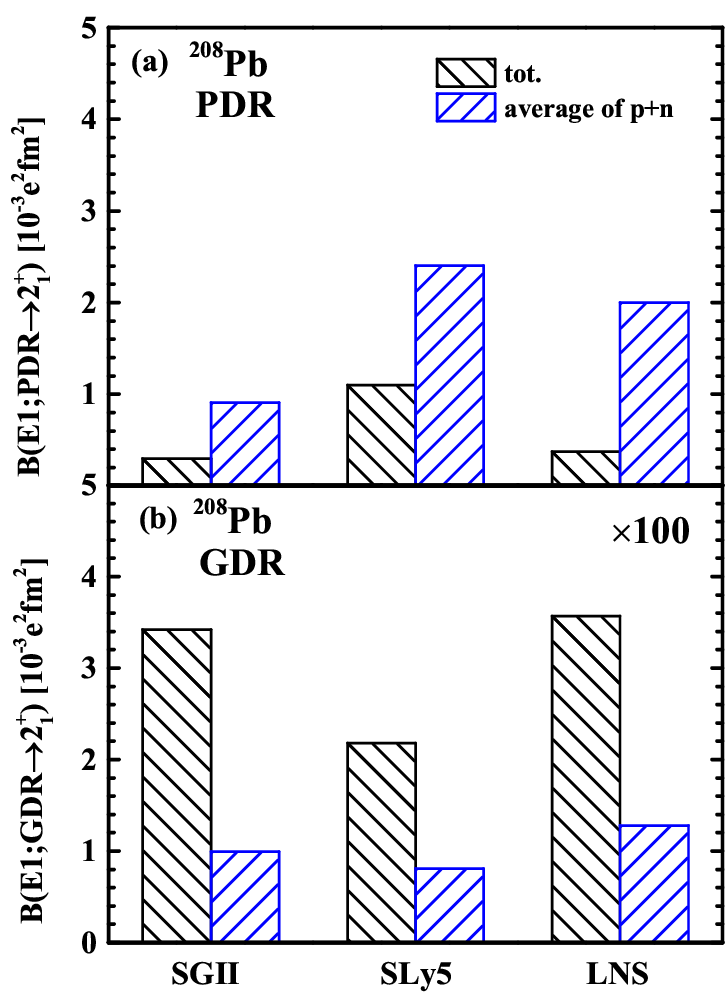}
\caption{The $\gamma$ transition strengths of PDR (a) and GDR (b) to $2_{1}^{+}$ in $^{208}$Pb for the 3 Skyrme interactions considered in this work.
        The black and blue hatched bars represent the total transition strength $B_{\gamma}$ and
        the average of proton and neutron contributions $\bar{B}_{\gamma}^{pn}$, respectively.}
\label{fig:fig3}
\end{figure}

To determine the isospin character of the selected pygmy dipole states, 
we analyze their $\gamma$-decay to the $2_1^{+}$ state.
Since the GDR at 13.5 MeV is a well-known isovector excitation, it can serve as a reference; 
on the other hand, 
the $2_1^{+}$ state is of isoscalar nature.
The $E1$ operator is isovector and, if isospin were an exact
quantum number, the $E1$ transition between $T=0$ states would be strictly forbidden. In a microscopic calculation
on a single-particle basis, the $E1$ operator gives opposite weights to protons and neutrons and, if the initial and final states
are isoscalar, there will be a cancellation between proton and neutron transition amplitudes that leads to a vanishing 
transition strength. Isospin is, in general, an approximate quantum number; in realistic calculations, 
isospin is physically broken by the Coulomb interaction and may be also broken within a specific many-body approximation.
We had previously checked, however, that in the framework of our model and recovering exact isospin symmetry 
we recover the $E1$ selection rule~\cite{Lv2021}.

We therefore compare the transition strengths of 
${\rm PDR} \rightarrow 2_1^+$ and ${\rm GDR} \rightarrow 2_1^+$.
Figure \ref{fig:fig3} shows the corresponding transition strengths $B_{\gamma}$ 
and the average of proton and neutron contributions $\bar{B}_{\gamma}^{pn}$.
For the PDR decay, $B_{\gamma}$ is less than half of $\bar{B}_{\gamma}^{pn}$; 
for the GDR decay, $B_{\gamma}$ exceeds twice $\bar{B}_{\gamma}^{pn}$.
This clear contrast indicates that the selected pygmy states exhibit an isoscalar character, 
while the GDR, as expected, retains its strong isovector nature.


\begin{figure}[t]
\centering
\includegraphics[width=0.8\linewidth]{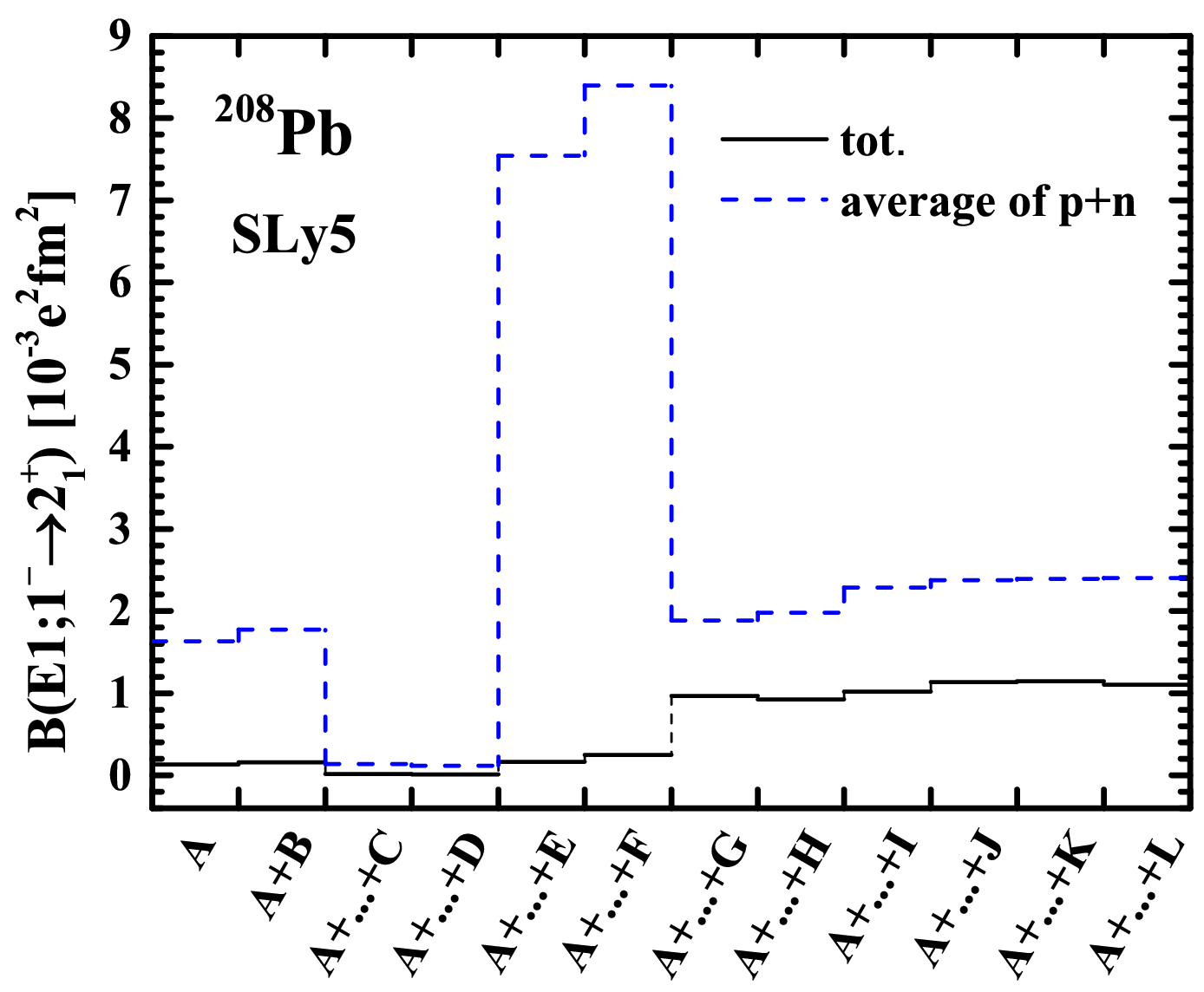}
\caption{Cumulative $\gamma$ transition strengths of the 12 diagrams of the selected pygmy dipole states
        to $2_{1}^{+}$ in $^{208}$Pb with the SLy5 interaction.
        The black solid and blue dashed lines represent the total transition strength $B_\gamma$ and
        the average of proton and neutron contributions $\bar{B}_{\gamma}^{pn}$, respectively.}
\label{fig:fig4}
\end{figure}

We further analyze the contribution from each diagram 
to deduce the information of the microscopic wave function. 
In Fig. \ref{fig:fig4}, we plot the cumulative sum of 
the transition strength $B_{\gamma}$ for the 12 diagrams calculated using the SLy5 functional.
Note that results of SGII and LNS are very similar.
To exclude the influence from the isospin character of PDR, 
the average proton-neutron contributions $\bar{B}_{\gamma}^{pn}$ is also shown.
For the cumulative sum of $B_{\gamma}$, the dominant diagram is G,
which corresponds to the transition between hole states.
This indicates that the 1p-1h configuration coupled to low-lying $2_{1}^{+}$ phonon (1p-1h$\otimes 2_{1}^{+}$)
constitutes a significant component of the PDR wave function.
More details emerge from the cumulative sum of $\bar{B}_{\gamma}^{pn}$.
First, a remarkable cancellation appears
between particle-type and hole-type transition contributions. 
This behavior is expected on general grounds, 
since particle- and hole-type contributions enter the transition amplitude with opposite phases 
and therefore interfere destructively \cite{Bortignon1981,Bortignon1984}.
At the RPA level, this cancellation occurs between diagrams (A,B) and (C,D). 
A similar cancellation persists at the PVC level between diagrams (E,F) and (G,H). 
Second, because of the strong cancellation at the RPA level, 
the net value of $\bar{B}_{\gamma}^{pn}$ is dominated by the PVC diagrams, particularly E and G.
This reveals the essential role of complex configurations in the PDR wave function,
namely the coupling between single-particle degrees of freedom and the $2_{1}^{+}$ phonon. 
Third, comparing the cumulative sums of $B_{\gamma}$ and $\bar{B}_{\gamma}^{pn}$ 
shows a partial cancellation between proton and neutron contributions in each diagram. 
This behavior confirms the predominant isoscalar character of the PDR.
Overall, both $B_{\gamma}$ and $\bar{B}_{\gamma}^{pn}$
are primarily determined by 
the component of 1p-1h$\otimes 2_{1}^{+}$ in the PDR wave function.
This conclusion, regarding the nature of the PDR wave function, 
is consistent with the analysis of Ref.~\cite{Spieker2020}. 

\begin{table}[b]
    \centering
    \caption{
        The average of the $\gamma$-decay strengths from proton and neutron, $\bar{B}_{\gamma}^{pn}$,
        The contribution from diagram E - H, $\bar{B}^{pn}_{\gamma~\rm E-H}$,
        as well as their relative ratios $R = \bar{B}^{pn}_{\gamma~\rm E-H}/\bar{B}_{\gamma}^{pn}$. 
        Units of $B$ are in $10^{-3} e^2 {\rm fm}^2$. 
        See the main text for a discussion related to the meaning of these quantities.
        }
    \label{tab:pvc_cont}
{\renewcommand{\arraystretch}{1.2}
 \begin{tabular}{c w{c}{1.5cm} w{c}{1.5cm} w{c}{1.5cm} w{c}{1.5cm} }
\hline \hline 
        &                             EDF   &   SGII    &  SLy5   &  LNS     \\  \hline
PDR     & $\bar{B}^{pn}_{\gamma}$           &   0.91    &  2.40   &  2.00    \\ 
        & $\bar{B}^{pn}_{\gamma~\rm E-H}$   &   0.55    &  1.15   &  0.67    \\ 
        & $R$                               &   61\%    &  48\%   &  33\%    \\  \hline
GDR     & $\bar{B}^{pn}_{\gamma}$           &   99.3    &  81.1   &  127.7   \\ 
        & $\bar{B}^{pn}_{\gamma~\rm E-H}$   &   63.6    &  76.0   &  101.4   \\ 
        & $R$                               &   64\%    &  94\%   &  79\%    \\  \hline
GQR     & $\bar{B}^{pn}_{\gamma}$           &  572.2    & 602.6   &  439.9   \\ 
        & $\bar{B}^{pn}_{\gamma~\rm E-H}$   &  564.0    & 574.3   &  444.7   \\ 
        & $R$                               &   99\%    &  95\%   &  101\%   \\  
\hline \hline 
\end{tabular}}
\end{table}

The $\gamma$-decay strength $B_{\gamma}$ from the PDR to the $2_1^{+}$ state is influenced
by several factors, including its isospin character, 
the contribution from complex configurations in its wave function,
and the collectivity of the resonance. Therefore, to isolate and quantitatively investigate the component of 1p-1h$\otimes 2_{1}^{+}$,
we proceed as follows.
We first calculate the average proton-neutron transition strength $\bar{B}_{\gamma}^{pn}$,
which effectively removes the isospin effects.
To specifically probe the role of the low-lying $2_{1}^{+}$ phonon,
we then calculate $\bar{B}_{\gamma~\rm E-H}^{pn}$, where only diagrams E-H are considered,
and compare it with the corresponding values for the GDR and GQR.
However, the absolute magnitude of $\bar{B}_{\gamma~\rm E-H}^{pn}$ 
is still affected by the collectivity of each resonance. 
To factor out this influence, we consider the ratio
$R=\bar{B}_{\gamma~\rm E-H}^{pn}/\bar{B}_{\gamma}^{pn}$,
which reflects the proportion of the 1p-1h$\otimes 2_{1}^{+}$ contribution 
in the total proton-neutron average strength.
The results, listed in Tab. \ref{tab:pvc_cont}, 
show a consistent hierarchy across all Skyrme EDFs employed,
$R_{\rm PDR} < R_{\rm GDR} < R_{\rm GQR}$.
This implies that the amount of the 1p-1h$\otimes 2_{1}^{+}$ component in the PDR is smaller than that in the GDR,
and much smaller than that in the GQR.
We note that the observed relation $R_{\rm GDR} < R_{\rm GQR}$ 
is in agreement with our previous theoretical studies \cite{Lv2021},
as well as with the wavelet analyses of experimental spectra \cite{Shevchenko2004, Poltoratska2014}.

\section{Summary}
\label{secSum}

In summary, the $\gamma$-decay of the pygmy dipole resonance 
to the $2_{1}^{+}$ state in $^{208}$Pb has been analyzed with
the Skyrme PVC model
to elucidate its isospin character and its microscopic structure.
The PDR states, located near the neutron emission threshold, 
show a pronounced cancellation between proton and neutron contributions 
to the $E1$ decay strength $B_{\gamma}$.
This cancellation signals a predominantly isoscalar character, 
distinct from the constructive coherence of proton and neutron in the case of the isovector GDR.
In simpler terms, since the 2$^+_1$ state is mainly $T=0$, the IVGDR has a strong $E1$ decay because it
is mainly $T=1$ and the weak $E1$ decay of the PDR points to its $T=0$ character.

By analyzing the cumulative sum from individual diagrams, 
we infer that the $\gamma$ decay strength is primarily determined by
the 1p-1h$\otimes 2_{1}^{+}$ configuration in the PDR wave function.
To isolate this phonon-coupled contribution from isospin effects and collectivity of resonance, 
we introduce the ratio $R=\bar{B}_{\gamma~\rm E-H}^{pn}/\bar{B}_{\gamma}^{pn}$. 
The systematic hierarchy $R_{\rm PDR} < R_{\rm GDR} < R_{\rm GQR}$ 
demonstrates quantitatively that the admixture of the low-lying $2_1^+$ phonon in the PDR 
is smaller than that in the GDR and much smaller than that in the GQR.

These findings confirm that the $\gamma$ decay to low-lying collective states 
provides a unique and sensitive observable for probing the isospin properties 
and the admixture of complex configurations in the pygmy dipole resonances.

\section*{Acknowledgements}
W.L.L. acknowledges helpful discussions with Dr. Yi-Wei Hao and Chen Chen.
This work was supported by 
the ``Young Scientist Scheme'' of National Key Research and Development (R\&D) Program under grant No. 2021YFA1601500,
the National Natural Science Foundation of China under grant Nos. 12405135, 12447168, 12075104, 12447106, 
the Science and Technology Innovation Leading Talent Project of Gansu Province (25RCKA025), 
the Lingchuang Research Project of China National Nuclear Corporation (CNNC-LCKY-2024-082), 
the Fundamental Research Funds for the Central Universities (lzujbky-2023-stlt01).

\bibliography{REFs_pygmy}

\end{document}